\documentclass[aps,prd,preprint,superscriptaddress,nofootinbib,amsmath,amssymb,showpacs]{revtex4}
\usepackage{amsmath}
\usepackage{bm}

\newcommand{\gtsim}{\protect\raisebox{-0.7ex}{$\:\stackrel{\textstyle >}{\sim}\:$}} 

\begin{document}
\preprint{RESCEU-40/12}
\title{Higgs condensation as an unwanted curvaton}


\author{Taro Kunimitsu}
\email[Email: ]{kunimitsu"at"resceu.s.u-tokyo.ac.jp}
\affiliation{Department of Physics, Graduate School of Science,
The University of Tokyo, Tokyo 113-0033, Japan}
\affiliation{Research Center for the Early Universe (RESCEU), Graduate
School of Science, The University of Tokyo, Tokyo 113-0033, Japan}
\author{Jun'ichi~Yokoyama}
\email[Email: ]{yokoyama"at"resceu.s.u-tokyo.ac.jp}
\affiliation{Research Center for the Early Universe (RESCEU), Graduate
School of Science, The University of Tokyo, Tokyo 113-0033, Japan}
\affiliation{Kavli Institute for the Physics and Mathematics of the Universe
(IPMU), The University of Tokyo, Kashiwa, Chiba, 277-8568, Japan}


\date{\today}

\begin{abstract}
\baselineskip 0.5cm
During inflation in the early universe, the Higgs field continuously acquires long-wave quantum fluctuations. They accumulate to yield a non-vanishing value with an exponentially large correlation length. We study consequences of such Higgs condensations to show that, in inflation models where the universe is reheated through gravitational particle production at the transition to the kination regime, they not only contribute to reheat the universe but also act as a curvaton. Unfortunately, however, for parameters of the Standard Model Higgs field, this curvaton produces density fluctuations too large, so the inflation models followed by a long kination regime are ruled out. 
\end{abstract}


\maketitle
\baselineskip 0.5cm
\section{Introduction}

The recent discovery of the new Higgs-like scalar particle by ATLAS and CMS at the Large Hadron Collider (LHC) \cite{LHC} is likely a great step toward confirming the Standard Model (SM) of elementary particle physics. The discovery of the spinless particle has a profound implication for cosmology, whether it will actually prove to be the SM Higgs field or not. This is especially true for inflation in the early universe \cite{inflation}, which is an indispensable ingredient of modern cosmology to solve the horizon and flatness problems and to account for the origin of density fluctuations \cite{fluctuation}, since the existence of a scalar field is required to drive inflation.

In this respect, the recent discovery at LHC may be directly connected with the Higgs inflation models \cite{higgsinflation}, where the SM Higgs field plays the role of the inflaton, the scalar field responsible for inflation. The SM Higgs field, however, cannot act as the inflaton adequately without extensions such as introducing a large and negative non-minimal coupling to the scalar curvature \cite{higgsinflation}, a Galileon like coupling \cite{higgsg}, a kinetic coupling to the Einstein tensor \cite{nonder}, a non-canonical kinetic term \cite{nakayama_takahashi}, etc. In fact, all the five known variants of the Higgs inflation can be treated in a unified way in the context of the generalized G-inflation model \cite{G2inflation}, which is the most general single-field inflation model with second-order field equations.

Even if inflation is induced by a scalar field of another sector in the full theory, the Higgs field must exist in the SM in any case, and it plays non-negligible roles in the evolution of the early universe. Indeed, a renormalization group analysis shows that the self coupling parameter $\lambda (\mu)$ in the Higgs potential may become negative at some large renormalization scale $\mu$, depending on the values of the Higgs mass $m_h$, top quark mass $m_t$, and the strong coupling constant $\alpha_S (M_Z)$. If the Higgs field acquires an expectation value larger than this critical value, it would exhibit a runaway behavior to a large and negative potential energy which would not only hamper inflation, but would also forbid the standard big bang cosmology \cite{Espinosa:2007qp}.

In December 2011, some hints of experimental signatures of the SM Higgs field were reported by ATLAS and CMS, with its preferred mass range $115 {\mathrm{GeV}} < m_h < 131 {\mathrm{ GeV}}$ by ATLAS and $m_h < 127 {\mathrm{GeV}}$ by CMS. Motivated by this result, a precise stability analysis of the Higgs vacua was done in \cite{running}, in the range $124 {\mathrm{GeV}} \leq m_h \leq 126 {\mathrm{GeV}}$; the study found that for $m_h = 124 {\mathrm{GeV}}$, the SM vacuum is at best metastable with a sufficiently long lifetime, and that the potential develops an instability at field value $10^{9-14} {\mathrm{GeV}}$, depending on the values of $m_t$ and $\alpha_S$. With a larger value of the Higgs mass, $m_h = 126 {\mathrm{GeV}}$, they find a finite parameter space of $m_t$ and $\alpha_S$ within $2\sigma$ where the SM vacuum is stable, with $\lambda (\mu)$ being positive for $\mu$ at least up to the Planck scale. Even in the case where the vacuum is metastable, the scale where $\lambda(\mu)$ becomes negative is pushed to a much higher scale than in the case with $m_h = 124{\mathrm{GeV}}$. Therefore, it was really fortunate for cosmology that the two collaborations pinned down the mass at a higher value $m_h \approx  126 {\mathrm{GeV}}$, allowing more room for the Higgs field value without any runaway behavior.

In this paper, we study cosmological consequences of such a Higgs field, assuming that the self coupling remains positive up to a field value well above the scale of inflation. During inflation, the Higgs field acquires a nonvanishing value by accumulating quantum fluctuations, a process which we will call the Higgs condensation. The fluctuations are exponentially stretched by the subsequent inflation, and its long-wave modes become indistinguishable from a homogeneous mode. As a result, the universe will be filled with many exponentially large coherent domains, each with different field values.

The typical value and correlation length of the condensate can be calculated using the stochastic inflation method \cite{stochastic1}. We find that for a natural value of the self coupling at the scale of inflation, the Higgs field acquires a fairly large value with an exponentially large coherent length compared to the Hubble radius during and just after inflation. Thus, when we calculate its effect on the reheating process, we can treat it as a homogeneous field, which starts coherent oscillation as its effective mass surpasses the Hubble parameter. On the other hand, the Higgs field is by no means fully homogeneous, so in the case where its condensate or its decay product significantly contributes to the total energy density of the universe, its fluctuation can also contribute to curvature perturbations through the curvaton mechanism \cite{curvaton}. We show that this is indeed the case for k-inflation \cite{k-inflation} and kinetically driven G-inflation \cite{G-inflation}, in which the universe is reheated through gravitational particle production. Our result would also apply to the quintessential inflation \cite{PeeblesVilenkin}, although it is difficult to realize this with a simple polynomial potential.

The rest of the paper is organized as follows. In Sec. II, we calculate the basic properties of the Higgs condensation using the stochastic inflation method. In Sec. III, gravitational reheating after k-inflation is calculated using standard techniques. In Sec. IV, the contribution of the Higgs field fluctuation on curvature perturbations is discussed.\footnote{The possibility of the Higgs field acting as a curvaton in an inflation model based on asymptotically safe gravity has been discussed in \cite{CaiEasson}.  A different possibility in which the Higgs field affects the density perturbation has recently been analyzed in \cite{DeSimoneRiotto}, where the Higgs field is always a minor component of the cosmic energy density.}

\section{Higgs condensation in inflationary cosmology}

We focus on the real neutral component of the SM Higgs field and denote it by $\varphi(\bm{x}, t)$. At the energy scale of inflation and reheating, which we assume is much larger than the electroweak scale, the potential of the Higgs field is well approximated by
\begin{equation}V(\varphi) = \frac{\lambda}{4} \varphi^4.\end{equation}

The self copuling $\lambda \equiv \lambda (\mu)$ has a logarithmic dependence on the energy scale (or field value). For the typical scale of inflation, we expect $\lambda (\mu) = {\mathcal O}(10^{-2})$ \cite{running, HH}.
Here we assume that $\lambda$ is constant, and take its reference value as $\lambda= 0.01$ hereafter. We also assume that the Higgs field is minimally coupled to gravity \cite{Spokoiny}.

The behavior of such a scalar field during inflation has been adequately studied in \cite{stochastic2}, using the stochastic inflation method, according to which the one-point probability distribution function (PDF) of $\varphi$, denoted by $\rho_1 (\varphi, t)$, is determined by the Fokker-Planck equation,
\begin{equation}
\frac{\partial \rho_1[\varphi({\bf x}, t)]}{\partial t} = \frac{1}{3H}\frac{\partial}{\partial \varphi} \{V^\prime[\varphi({\bf x}, t)]\rho_1[\varphi({\bf x}, t)]\} + \frac{H^3}{8\pi^2}\frac{\partial^2 \rho_1[\varphi({\bf x}, t)]}{\partial \varphi^2} \equiv \Gamma_\varphi \rho_1 [\varphi({\bf x}, t)].
\end{equation} 
Its generic solution can be expanded as
\begin{align}
\rho_1(\varphi, t) &= \exp \left(-\frac{4\pi^2V(\varphi)}{3H^4}\right)\sum^{\infty}_{n=0} a_n \Phi_n (\varphi) e^{-\Lambda_n (t-t_0)}\\ &= \rho_{1\mathrm{eq}}(\varphi) +  \exp \left(-\frac{4\pi^2V(\varphi)}{3H^4}\right)\sum^{\infty}_{n=1} a_n \Phi_n (\varphi) e^{-\Lambda_n (t-t_0)},\end{align}
where $\Phi_n(\varphi)$ is the complete orthonormal set of eigenfunctions of the Schr$\ddot{\mathrm{o}}$dinger-type equation,
\begin{equation}\left[-\frac{1}{2}\frac{\partial^2}{\partial \varphi^2} + W(\varphi) \right] \Phi_n(\varphi) =  \frac{4\pi^2 \Lambda_n}{H^3} \Phi_n(\varphi),\end{equation}
with
\begin{equation}W(\varphi) \equiv \frac{1}{2}\left[v^\prime (\varphi) ^2 - v^{\prime\prime}(\varphi)\right], \label{Schroedinger}\end{equation}
and
\begin{equation}v(\varphi) \equiv \frac{4\pi^2}{3H^4}V(\varphi).\end{equation}
For $V(\varphi) = \frac{\lambda}{4}\varphi^4$ we find the equilibrium one-point PDF
\begin{equation}\rho_{1\mathrm{eq}} (\varphi) = \left( \frac{32\pi^2\lambda}{3}\right)^{\frac{1}{4}} \frac{1}{\Gamma(\frac{1}{4})H} \exp \left(-\frac{2\pi^2 \lambda \varphi^4}{3H^4}\right),\end{equation}
as well as the first few eigenvalues of (\ref{Schroedinger}) as,
\begin{equation}\Lambda_0 =0, \ \ \ \ \ \Lambda_1 = 1.36859 \sqrt{\frac{\lambda}{24\pi^2}}H, \ \ \ \  \ \Lambda_2 = 4.4537\sqrt{\frac{\lambda}{24\pi^2}}H,  \label{eigenvalue}\end{equation}
numerically \cite{stochastic2}, which are useful in evaluating correlation functions.

As a result, we find the following equilibrium expectation values for each quantity,
\begin{equation}\langle \varphi^2\rangle = \sqrt{\frac{3}{2\pi^2}}\frac{\Gamma(\frac{3}{4})}{\Gamma(\frac{1}{4})} \frac{H^2}{\sqrt{\lambda}} \simeq 0.132\frac{H^2}{\sqrt{\lambda}} =1.32{\tilde{\lambda}}^{-\frac{1}{2}} H^2,\end{equation}
\begin{equation}m_{\mathrm{eff}}^2 \equiv \lambda \langle \varphi^2\rangle \simeq 1.32\times 10^{-2} {\tilde{\lambda}}^{\frac{1}{2}}H^2,\label{meff}\end{equation}
\begin{equation}\langle V(\varphi) \rangle = \frac{\lambda}{4} \langle \varphi^4 \rangle= \frac{3H^4}{32\pi^2} \simeq 9.50 \times 10^{-3} H^4,\end{equation}
where we have defined ${\tilde{\lambda}}\equiv 10^2 \lambda \sim 1$.

Using the eigenvalues (\ref{eigenvalue}), in particular the fact that $\Lambda_2$ is significantly larger than $\Lambda_1$, the temporal auto correlation function is well approximated by
\begin{equation}G(t_1 - t_2) \equiv \langle \varphi({\bf x}, t_1)\varphi({\bf x}, t_2) \rangle \simeq \langle \varphi^2 \rangle e^{-\Lambda_1|t_1 - t_2|}, \label{temporal}\end{equation}
for $|t_1-t_2| \gtsim H^{-1}$.
The correlation time, $t_c$, which is defined by $G(t_c) = \frac{1}{2} G(0)$, is given by $t_c \simeq 76.2 {\tilde{\lambda}}^{-\frac{1}{2}} H^{-1}$. 

The spatial correlation function can also be evaluated making use of its de Sitter invariance \cite{stochastic2}. Replacing $t_1 - t_2$ by $\frac{2}{H}\ln(Har)$ in (\ref{temporal}), we find
\begin{equation}G(r) \equiv \langle \varphi({\bf x}_1, t)\varphi({\bf x}_2, t) \rangle \simeq \langle \varphi^2 \rangle (H a(t) r )^{-\frac{2\Lambda_1}{H}}, \label{spatial}\end{equation} 
where $r \equiv |{\bf x}_1 - {\bf x}_2|$. The spatial correlation length is defined by $G(r_c) = \frac{1}{2} G(0)$, and reads
\begin{equation}a(t) r_c = \frac{H^{-1}}{2} e^{\frac{H}{2\Lambda_1}} \simeq \frac{H^{-1}}{2} e^{38.1 {\tilde{\lambda}}^{-\frac{1}{2}}}, \end{equation}
which is exponentially larger than the Hubble radius during inflation, as well as that at the end of reheating, if not larger than the comoving scale corresponding to the current horizon. Hence one can regard the field configuration to be practically homogeneous when we discuss the reheating processes.

This means that all the massive fermions and gauge bosons in the standard model acquire a mass squared proportional to $\langle \varphi^2 \rangle$. For quarks and leptons the resultant masses are substantially smaller than the Hubble parameter (except for the top quark) because they are suppressed by the Yukawa coupling. 

On the other hand, gauge bosons could have a mass close to the Hubble parameter. It breaks the conformal invariance of the vector field, and each species of gauge bosons may be created gravitationally with an amount close to the Higgs boson. This issue will be discussed elsewhere. Here we incorporate it in the effective number of modes of created particles, N, to be defined below.

\section{Gravitational reheating after k-inflation}

In this paper we consider consequences of the Higgs condensation discussed in the previous section, in inflation models where the universe is reheated by gravitational particle production without being accompanied by the inflaton's field oscillation \cite{k-inflation, G-inflation, PeeblesVilenkin}.

Specifically, let us consider k-inflation \cite{k-inflation} with its Lagrangian of the form
\begin{equation}
{\mathcal L} = K_1(\phi) X + K_2(\phi) X^2, \ \ \ X\equiv -\frac{1}{2}g^{\mu\nu}\partial_\mu\phi\partial_\nu\phi,
\end{equation}
where $\phi$ is the inflaton and $X$ is its canonical kinetic function. The fact that such a theory can realize inflation even without any potential term can most easily be seen by considering the situation where $K_1$ and $K_2$ are constants with opposite sign. Then in the homogeneous and isotropic background, X has an attractor solution $X= - \frac{K_1}{2K_2} > 0$. The cosmic energy density, $\rho$, and the pressure, $P$, read
\begin{equation}
\rho = 2X\frac{\partial{\mathcal{L}}}{\partial X} -{\mathcal{L}} = -P = \mathrm{constant},
\end{equation}
which induces an exponential cosmic expansion with the Hubble parameter squared $H^2_{\mathrm{inf}}= \frac{2\pi G}{3}\frac{K_1^2}{K_2}$.

In this model, inflation terminates when $K_1$ and $K_2$ both become positive. Then, the kinetic energy starts to redshift quickly and only the first term in the Lagrangian becomes relevant. That is, the universe is dominated by the kinetic energy of a free scalar field with its equation of state $w\equiv P/\rho =1$.
Then, the scale factor $a(t)$ scales as $a(t) \propto t^{1/3}$, or in terms of conformal time, $\eta$, it asymptotically behaves as $a(t) \propto \eta^{1/2}$. The energy density decreases as $a^{-6}$, and the scalar curvature is given by $R=-6H^2 (t)$.

If we assume that the universe underwent an immediate transition from the de Sitter phase to the kinetic energy dominated phase (kination phase), the scale factor right after inflation can be written using conformal time $\eta$ as
\begin{equation}a^2(\eta) = \frac{2}{H_{\mathrm{inf}}^2|\eta_0|^3}(\eta - \eta_0) + \frac{1}{H_{\mathrm{inf}}^2|\eta_0|^2}. \label{a2}\end{equation}
$H_{\mathrm{inf}}$ here is the Hubble constant during inflation and $\eta_0<0$ is the value of the conformal time at the end of inflation. We will take $a(\eta_0)=\frac{1}{H_{\mathrm{inf}}|\eta_0|}$ to be equal to 1 by adjusting $\eta_0$, which gives
\begin{equation}a^2(\eta) = \frac{2}{|\eta_0|}(\eta - \eta_0) + 1. \end{equation}

Reheating in k-inflation is realized by gravitational particle production, due to the change in the definition of the vacuum. 
Following the standard calculation of particle creation \cite{particlecreation, Ford}, we consider the creation of massless minimally coupled scalar particles. We can calculate the Bogolubov coefficient $\beta_\omega$,
\begin{equation}\beta_\omega =\frac{i}{2\omega}\int^{\infty}_{-\infty} e^{-2i\omega\eta}V(\eta)d\eta, \end{equation}
where $V(\eta) = \frac{1}{6} a^2(\eta) R(\eta)$, and we attain expressions for the number density and energy density
\begin{equation}n_r= \frac{1}{2\pi^2 a^3}\int^{\infty}_{0}|\beta_\omega|^2\omega^2d\omega, \end{equation}
\begin{equation}\rho_r = \frac{1}{2\pi^2 a^4}\int^{\infty}_{0} |\beta_\omega|^2 \omega^3d\omega. \end{equation}
The $r$ in $n_r$, $\rho_r$ stands for the relativistic particles created at reheating. 

During inflation, $\beta_\omega$ is proportional to $\omega^{-2}$ at high energies, so the integral for $\rho_r$ diverges logarithmically. This divergence is due to the discontinuity of $R$, which can be seen by rewriting $\rho_r$ as
\begin{equation}\rho_r = -\frac{1}{32\pi^2a^4} \int^{\eta_0}_{-\infty} d\eta_1 \int^{\eta_0}_{-\infty} d\eta_2 \ln (|\eta_1 - \eta_2|\mu) V^{\prime}(\eta_1) V^{\prime}(\eta_2), \end{equation}
where  the primes denote the derivative with respect to $\eta$. $\mu$ is some arbitrary mass inserted for dimensional reasons ($\rho_r$ is independent of the value of $\mu$).

The above-mentioned divergence is the result of the sudden-transition approximation from the de Sitter phase $a(\eta) = -\frac{1}{H_{\mathrm{inf}}\eta}$ to (\ref{a2}). In order to obtain a finite value for the energy density, we consider the case in which the universe makes a transition from de Sitter space to a kinetic energy dominated universe within a timescale $\Delta \eta = H_{\mathrm{inf}}^{-1}x_0$ and keeping the Ricci scalar continuous, following Ford \cite{Ford} who actually considered a smooth transition from de Sitter to a power-law expansion with $a(t) \propto t^{1/2}$ instead of $t^{1/3}$. Since $a(\eta_0) =1$, we have $\Delta \eta \simeq \Delta t$. We will show that for an arbitrarily small but finite $\Delta t$,  the energy density after inflation can be analytically calculated.

Let us define 
\begin{equation}   f(H_{\mathrm{inf}} \eta) \equiv a^2 (\eta), \end{equation}
and rewrite the energy density as
\begin{equation}\rho_r = \frac{H_{\mathrm{inf}}^4}{128\pi^2 a^4} I,\end{equation}
where
\begin{equation}I = -\int^{x}_{-\infty} dx_1 \int^{x}_{-\infty} dx_2 \ln (|x_1 - x_2|) \tilde V^{\prime}(x_1) \tilde V^{\prime}(x_2),\end{equation}
\begin{equation}\tilde V(x) = \frac{f^{\prime\prime}f - \frac{1}{2}(f^{\prime})^2}{f^2}.\end{equation}
$x=H_{\mathrm{inf}} \eta$ is the time when $\tilde V$ becomes sufficiently smaller than $1$, at which the notion of particle is well-defined.

We make the following ansatz,
\begin{equation}
f(x)= \begin{cases}
\displaystyle \frac{1}{x^2} & (x<-1)\\
a_0 + a_1 x + a_2 x^2 + a_3 x^3 & (-1<x<x_0-1)\\
b_0 (x+b_1) & (x_0-1<x)
\end{cases}\end{equation}
and require $f(x), f^{\prime}(x), f^{\prime \prime}(x)$ to be continuous at $x=-1$ and $x=x_0 -1$ (which makes the Ricci scalar continuous throughout the period under consideration). If the transition from $K_1(\phi)K_2(\phi)\simeq \mathrm{constant} <0$ to $K_1(\phi)K_2(\phi)>0$ occurs in a time scale much less than the Hubble time, numerical calculations show that the transition to the kination regime with $w=1$ also takes place well within the Hubble time \cite{G-inflation}. Thus we can take $x_0<1$, or $\Delta t < H_{\mathrm{inf}}^{-1}$.

The coefficients are determined as (up to $O((x_0)^0)$),
\begin{align}
a_0 &= 6 - \frac{1}{x_0},& 
a_1 &= 8 - \frac{3}{x_0},&
a_2 &= 3 - \frac{3}{x_0},\\
a_3 &= -\frac{1}{x_0},&
b_0 &= 2,&
b_1 &= \frac{3}{2},
\end{align}
which leads to
\begin{equation}\tilde{V}^{\prime}=-\frac{6}{x_0}.\end{equation} 
We can see that $b_0$ and $b_1$ coincide with the expression for the scale factor (\ref{a2}), which justifies our ansatz.

We can now calculate $I$, which becomes
\begin{equation}I\sim -36\ln x_0, \end{equation}
so the energy density at the beginning of the kinetic energy dominated era is
\begin{equation}\rho_r = \frac{9 H_{\mathrm{inf}}^4}{32\pi^2 a^4} \ln \left(\frac{1}{H_{\mathrm{inf}}\Delta t}\right), \label{density}\end{equation}
This value is $\frac{9}{4}$ times larger than the value for the radiation dominated universe, derived in \cite{Ford}. 
We will ignore the dependence on $\Delta t$ which is logarithmic, and take $\ln(1/(H\Delta t))\sim 1$. If there are $N$ modes of species created this way, (\ref{density}) should be multiplied by $N$.

Now we go on to calculate the reheating temperature.
If we neglect the effect of Higgs condensation, the reheating temperature is determined by the condition $\rho_\phi = \rho_r$, where $\rho_\phi$ is the energy density of the inflaton field, 
\begin{equation}
\rho_\phi = 3 M_G^2 H^2_{\mathrm{inf}}a^{-6}. 
\end{equation}
Here $M_G$ is the reduced Planck mass. Then, the scale factor $a_R$ at the time of reheating satisfies
\begin{equation}
a_R^{2} =\frac{32 \pi^2M_G^2}{3 N H_{\mathrm{inf}}^2},
\end{equation}
so the energy density of the relativistic particles is
\begin{equation}
\rho_r|_R =  \frac{81 N^3 H_{\mathrm{inf}}^8}{(32\pi^2)^3 M_G^4}.
\end{equation}
In terms of the reheating temperature, the radiation energy density is expressed as $\displaystyle \frac{\pi^2 g_\ast}{30}T_R^4$, where $g_\ast$ is the effective number of relativistic species of particles, so
\begin{equation}T_R=\frac{3N^{\frac{3}{4}}}{(32\pi^2)^{\frac{3}{4}}}\left(\frac{30}{\pi^2 g_\ast}\right)^{\frac{1}{4}}\frac{H_{\mathrm{inf}}^2}{M_G} \simeq 3.9\times 10^6 N^{\frac{3}{4}}\left(\frac{g_\ast}{106.75}\right)^{-\frac{1}{4}}\left(\frac{r}{0.01}\right)\mathrm{GeV}.\end{equation}
Here $r$ denotes the tensor-to-scalar ratio which is given by 
\begin{equation}r=0.01\left(\frac{H_{\mathrm{inf}}}{2.4\times 10^{13} \mathrm{GeV}}\right)^2.\end{equation}

The above analysis applies in the case in which the Higgs condensation does not contribute to the cosmic energy density. Now we consider how the Higgs condensation affects the reheating temperature in this model. After inflation, the Higgs field remains constant until the Hubble parameter decreases below $m_{\mathrm{eff}} \simeq 0.115 {\tilde{\lambda}}^{1/4}H_{\mathrm{inf}}$ (see eq. (\ref{meff})). When $H$ becomes smaller than $m_{\mathrm{eff}}$, the Higgs field starts oscillating. At the beginning of the oscillation phase, the energy density of the inflaton, relativistic matter, and Higgs field can be expressed as
\begin{equation}
\rho_\phi = 3 M_G^2 m_{\mathrm{eff}}^2 \simeq 3.96 \tilde{\lambda} ^{\frac{1}{2}} M_G^2 H_{\mathrm{inf}}^2,
\end{equation}
\begin{equation}
\rho_r= \frac{9 N H_{\mathrm{inf}}^4}{32\pi^2}  \left(\frac{m_{\mathrm{eff}}}{H_{\mathrm{inf}}}\right)^{\frac{4}{3}} \simeq 1.59 \times 10^{-3} N  H_{\mathrm{inf}}^4,
\end{equation}
\begin{equation}
\rho_{\mathrm{cond}} = \frac{3  H_{\mathrm{inf}}^4}{32 \pi^2} \simeq 9.50 \times 10^{-3}  H_{\mathrm{inf}}^4,
\end{equation}
respectively. Thus $\rho_{\mathrm{cond}}$ can be larger than $\rho_r$ at this moment. Since both $\rho_r$ and $\rho_{\mathrm{cond}}$ scale as $a^{-4}$ after the oscillation begins \cite{Turner}, the universe can be predominantly reheated through the decay product of the Higgs condensation. In such a case, the reheating temperature can be estimated from the equality $\rho_\phi = \rho_{\mathrm{cond}}$, because the Higgs condensation dissipates its energy to radiation rapidly, once the field oscillation commences. As a result, we find
\begin{equation}T_R = 1.8\times 10^7 \left(\frac{g_\ast}{106.75}\right)^{-\frac{1}{4}}\left(\frac{r}{0.01}\right)\mathrm{GeV}.\end{equation}

\section{Fluctuation of the Higgs field after inflation}
So far we have treated the Higgs field as a practically homogeneous condensate after inflation, because it has an exponentially large correlation length compared to the horizon scale at the end of inflation. In reality, however, the Higgs field acquires long-wave quantum fluctuations whose observational consequence should be clarified. 

To do this, we first calculate its long-wave power spectrum from the spatial correlation function (\ref{spatial}). A power-law correlation function can be obtained from a power-law power spectrum $P(k) \equiv |\varphi_k|^2 \equiv Ak^n$ as,
\begin{equation}G(r) = \int{P(k) e^{-i{\bf k \cdot r}}} \frac{d^3 k}{(2\pi)^3}= Ar^{-n-3} \frac{\Gamma (n+2)}{2\pi^2} \sin \left[(n+2) \frac{\pi}{2}\right],\end{equation}
which, for the case we are considering, means $\displaystyle n=-3+\frac{2\Lambda_1}{H_{\mathrm{inf}}}$ and
\begin{equation}A\simeq \frac{4\pi^2 \Lambda_1}{H_{\mathrm{inf}}}\sqrt{\frac{3}{2\pi^2 \lambda}} \frac{\Gamma(\frac{3}{4})}{\Gamma(\frac{1}{4})} a^{-\frac{2\Lambda_1}{H}}H_{\mathrm{inf}}^{2-\frac{2\Lambda_1}{H}},\end{equation}
respectively. In deriving the latter, we have adopted an approximation that $\frac{2\Lambda_1}{H_{\mathrm{inf}}}$ is much smaller than unity.

As a result we find
\begin{equation}P(k) = |\varphi_k|^2 \simeq 0.462 \frac{H_{\mathrm{inf}}^2}{k^3} \left(\frac{k}{H}\right)^{\frac{2\Lambda_1}{H_{\mathrm{inf}}}} \approx \frac{H_{\mathrm{inf}}^2}{2k^3},\end{equation}
that is, its power-spectrum is very close to that of a massless minimally coupled field.

We can also calculate the power spectrum of the energy density fluctuation of the Higgs condensate after inflation. Since it is frozen while $m_{\mathrm{eff}}< H_{\mathrm{inf}}$, we can estimate it only from the potential energy fluctuation. Since the potential energy is also governed by the same PDF as used above, its large-scale correlation function is given by
\begin{align}
\Xi_{\Delta_h}(r) &\equiv \left\langle \frac{\delta \rho_h(r)}{\rho_h} \frac{\delta \rho_h(0)}{\rho_h} \right\rangle 
\cong  \frac{1}{\langle V \rangle ^2} \left[ \langle V(r) V(0) \rangle - \langle V(0) \rangle^2 \right]\notag\\
&\simeq  \frac{1}{\langle V \rangle ^2} \left[ \langle V^2\rangle - \langle V \rangle^2 \right] e^{-\Lambda_1 t_\ast}
=4(Har)^{-\frac{2\Lambda_1}{H}}\notag\\
&\equiv \int P_V(k) e^{-i{\bf k \cdot r} }\frac{d^3k}{(2\pi)^3},
\end{align}
where $t_\ast \equiv \frac{2}{H}\ln (Har)$.

The power spectrum $P_V(k)$ then reads $P_V(k) \approx 14\sqrt{\lambda}k^{-3}$, so the amplitude of density fluctuations on scale $r\equiv \frac{2\pi}{k}$ is given by
\begin{equation}{\mathcal P}_V(k) \equiv \frac{4\pi k^3}{(2\pi)^3} P_V(k) = 0.71 \sqrt{\lambda} = 0.071 {\tilde \lambda}^{\frac{1}{2}}\end{equation}
Thus the Higgs field acquires almost scale-invariant fractional density fluctuations with amplitude $\sim 0.1$.

Since this or its decay products will make a significant contribution to the energy density of the universe, if not dominant, after the inflaton's kinetic energy density has dissipated away according to $\rho_\phi \propto a^{-6}$, the Higgs field unexpectedly acts as a curvaton \cite{curvaton}. Its contribution to the final curvature perturbation is schematically given as \cite{KKT},
\begin{equation}
{\mathcal R} \approx \frac{\rho_h}{\rho_{\mathrm{tot}}}\left(c\frac{\delta \rho_h}{\rho_h} - c^\prime \frac{\delta H_{\mathrm{osc}}}{H_{\mathrm{osc}}}\right)\bigg|_{\mathrm{osc}} \sim \frac{\rho_h}{\rho_{\mathrm{tot}}}\bigg|_{\mathrm{osc}} {\mathcal P}^{\frac{1}{2}}_V (k),
\end{equation}
where $H_{\mathrm{osc}}$ is the Hubble parameter at the onset of Higgs oscillation ($=m_{\mathrm{eff}}$), $\delta H_{\mathrm{osc}}$ is its fluctuation, and $c$ and $c^\prime$ are quantities of order unity.
As a result, it induces curvature perturbations too large with the phenomenologically inferred value of $\lambda$, since $\frac{\rho_h}{\rho_{\mathrm{tot}}}$ at the onset of oscillation is ${\mathcal O}(1)$.

Thus the decay product of the Higgs condensation cannot constitute the dominant component of the universe. In other words, some part of the inflaton's energy must be directly transferred to radiation without too long a period of the kination regime.

\section{Discussion}
We have considered cosmology of the SM Higgs field based on the recent discovery of a Higgs like scalar particle with a mass $m_h \simeq 126 {\mathrm{GeV}}$. Assuming that its self coupling $\lambda(\mu)$ is positive and ${\mathcal{O}}(10^{-2})$ at the scale of inflation, we have analyzed its spatial configuration using the stochastic inflation method. We have found that for an experimentally inferred value of the self coupling $\lambda(\mu) \sim 0.01$ at the scale of inflation, the Higgs condensation due to long-wave quantum fluctuations acquired during inflation suffers from large density perturbations.

In the conventional potential-driven inflation, slow-roll inflation is followed by field oscillation, whose energy dissipates in proportion to $a^{-3}$ until reheating takes place. Hence the energy density of the Higgs condensation, which is much smaller than the inflaton's energy density with $\rho_{\mathrm{cond}} / \rho_\phi \simeq H_{\mathrm{inf}}^2/M_G^2 <10^{-5}$, never contributes to the total energy density appreciably, since its coherent oscillation dissipates in proportion to $a^{-4}$.

In k-inflation or kinetically driven G-inflation as well as in quintessential inflation, reheating occurs through gravitational particle production and the inflaton energy dissipates as $a^{-6}$. As a result, even if the oscillation of $\varphi$ is governed by the quartic potential and its energy density scales as $a^{-4}$, it will eventually contribute to the total energy density significantly and act as a curvaton. This is an unusual case, since a scalar field with such a quartic potential usually does not play the role of a curvaton as explained above (see also \cite{KKT, Enqvist:2009zf} for other effects of self interaction).

Thus, k-, G-, and quintessential inflation with long kination regimes are ruled out due to the SM Higgs field.  It has been argued that the imprint of the kination regime can be observed as an enhancement of the high-frequency part of the stochastic gravitational wave background energy density produced by inflation \cite{Tashiro:2003qp}. Our analysis shows that such a possibility is inconsistent with the observed amplitude of curvature perturbations, due to the extra contribution of the Higgs field.

There is, however, a simple remedy. If radiation particles are directly created from the inflaton due to some direct coupling, then reheating would be more efficient.  Another possibility is to extend the structure of the Higgs field.  For example, if one introduces supersymmetry, the condensate of the $H_uH_d$ flat direction may dominate the energy density of the post-inflationary universe eventually.  Then it not only works as a curvaton but also creates practically all the radiation components \cite{EKM}. In either case, a long kination regime would not be realized after inflation and the enhancement of the high frequency gravitational wave background would be impossible. Thus the observational significance of our analysis mentioned above does not change even with these solutions.

\section*{Acknowledgments}

We would like to thank Kohei Kamada for useful communications.
This work was partially supported by
JSPS Grant-in-Aid for Scientific Research
No. 23340058 (J.Y.), and the Grant-in-Aid for
Scientific Research on Innovative Areas No. 21111006 (J.Y.).


\begin{thebibliography}{99}
\bibitem{LHC}
ATLAS Collaboration [arXiv:1207.7214 [hep-ex]]; CMS Collaboration [arXiv:1207.7235 [hep-ex]]

\bibitem{inflation}
A. A. Starobinsky, Phys. Lett. B {\bf 91}, 99 (1980); 
K. Sato, Mon. Not. Roy. Astron. Soc. {\bf 195}, 467 (1981); 
A. H. Guth, Phys. Rev. D {\bf 23}, 347 (1981)

\bibitem{fluctuation}
V. F. Mukhanov, G. V. Chibisov, JETP Lett. {\bf 33}, 532-535 (1981); 
S. W. Hawking, Phys. Lett. B {\bf 115}, 295 (1982); 
A. A. Starobinsky, Phys. Lett. B {\bf 117}, 175 (1982); 
A. H. Guth and S-Y.~Pi, Phys. Rev. Lett. {\bf 49}, 1110 (1982)

\bibitem{higgsinflation}
J.~L.~Cervantes-Cota and H.~Dehnen,
 Nucl.\ Phys.\ B {\bf 442}, 391 (1995)
 [arXiv:astro-ph/9505069];
F. L. Bezrukov, M. Shaposhnikov, Phys. Lett. B {\bf 659}, 703 (2008) [arXiv:0710.3755[hep-th]];
  A.~O.~Barvinsky, A.~Y.~Kamenshchik and A.~A.~Starobinsky,
  JCAP {\bf 0811}, 021 (2008)
  [arXiv:0809.2104 [hep-ph]]; 
  F.~Bezrukov, D.~Gorbunov and M.~Shaposhnikov,
  JCAP {\bf 0906}, 029 (2009)
  [arXiv:0812.3622 [hep-ph]]; 

\bibitem{higgsg}
K. Kamada, T. Kobayashi, M. Yamaguchi, J. Yokoyama, Phys. Rev. D {\bf 83}, 083515 (2011) [arXiv:1012.4238 [astro-ph.CO]]

\bibitem{nonder}
  C.~Germani and A.~Kehagias,
  Phys.\ Rev.\ Lett.\  {\bf 105}, 011302 (2010)
  [arXiv:1003.2635 [hep-ph]]; 
  C.~Germani and A.~Kehagias,
  JCAP {\bf 1005}, 019 (2010)
  [Erratum-ibid.\  {\bf 1006}, E01 (2010)]
  [arXiv:1003.4285 [astro-ph.CO]]; 
 C.~Germani and Y.~Watanabe,
 JCAP {\bf 1107}, 031 (2011)
 [Addendum-ibid.\  {\bf 1107}, A01 (2011)]
 [arXiv:1106.0502 [astro-ph.CO]]
 
\bibitem{nakayama_takahashi}
K. Nakayama, F. Takahashi, JCAP {\bf 1102}, 010 (2011) [arXiv:1008.4457 [hep-ph]]


\bibitem{G2inflation}
T. Kobayashi, M. Yamaguchi, J. Yokoyama, Prog. Theor. Phys. {\bf 126}, 511 (2011) [arXiv:1105.5723 [hep-th]]

 \bibitem{Espinosa:2007qp}
  J.~R.~Espinosa, G.~F.~Giudice and A.~Riotto,
JCAP {\bf 0805}, 002 (2008)  [arXiv:0710.2484 [hep-ph]] 


\bibitem{running}
J. Elias-Mir$\mathrm{\acute{o}} $, J. R. Espinosa, G. F. Giudice, G. Isidori, A. Riotto and A. Strumia, Phys. Lett. B {\bf 709}, 222 (2012)
[arXiv:1112.3022[hep-ph]]

\bibitem{stochastic1}
A. A. Starobinsky, in {\it Fundamental Interactions}, edited by V. N. Ponomarev (MGPI Press, Moscow 1984), p.55;
A. A. Starobinsky, in {\it Field Theory, Quantum Gravity and Strings}, edited by H. J. de Vega and N. Sanchez, Lecture
Notes in Physics {\bf 246}, 107 (1986)

\bibitem{curvaton}
K. Enqvist and M. S. Sloth, Nucl. Phys. B {\bf 626}, 395 (2002) [arXiv:hep-ph/0109214];
D. H. Lyth and D. Wands, Phys. Lett. B {\bf 524}, 5 (2002) [arXiv:hep-ph/0110002];
T. Moroi and T. Takahashi, Phys. Lett. B {\bf 522}, 215 (2001) [Erratum-ibid. B 539,
303 (2002)] [arXiv:hep-ph/0110096]

\bibitem{k-inflation}
C. Armendariz-Picon, T. Damour and V. F. Mukhanov,
Phys. Lett. B {\bf 458}, 209 (1999) [arXiv:hep-th/9904075]


\bibitem{G-inflation}
T. Kobayashi, M. Yamaguchi, J. Yokoyama, Phys. Rev.
Lett. {\bf 105}, 231302 (2010) [arXiv:1008.0603 [hep-th]]

\bibitem{PeeblesVilenkin}
P. J. E. Peebles, A. Vilenkin, 
Phys. Rev. D {\bf 59} 063505 (1999) [arXiv:astro-ph/9810509] 

\bibitem{CaiEasson}
  Y.~-F.~Cai and D.~A.~Easson,
  arXiv:1202.1285 [hep-th]

\bibitem{DeSimoneRiotto}
A. De Simone, A. Riotto [arXiv:1208.1344 [hep-ph]]

\bibitem{HH}
J.A. Casas, J.R. Espinosa, M. Quiros, Phys. Lett. B {\bf 342}, 171 (1995);
J.A. Casas, J.R. Espinosa, M. Quiros, Phys. Lett. B {\bf 382}, 374 (1996);
G. Isidori, G. Ridolfi, A. Strumia, Nucl. Phys. B {\bf 609}, 387 (2001) [arXiv:hep-ph/0104016];
G. Isidori, V.S. Rychkov, A. Strumia, N. Tetradis, Phys. Rev. D {\bf 77}, 025034 (2008) [arXiv:0712.0242 [hep-ph]]

\bibitem{Spokoiny}
  B.~L.~Spokoiny,
  Phys.\ Lett.\  B {\bf 147}, 39 (1984);
  T.~Futamase and K.~-i.~Maeda,
  Phys.\ Rev.\ D {\bf 39}, 399 (1989);
  D.~S.~Salopek, J.~R.~Bond and J.~M.~Bardeen,
  Phys.\ Rev.\  D {\bf 40}, 1753 (1989); 
  R.~Fakir and W.~G.~Unruh,
  Phys.\ Rev.\  D {\bf 41}, 1783 (1990)


\bibitem{stochastic2}
A. A. Starobinsky and J. Yokoyama, Phys. Rev. D {\bf 50}, 6357 (1994) [arXiv:astro-ph/9407016]

\bibitem{particlecreation}
L. Parker, Phys. Rev. {\bf 183}, 1057 (1969);
Y. B. Zel'dovich, A. A. Starobinsky, Zh. Eksp. Toer. Fiz. {\bf 61}, 2161 (1971) [Sov. Phys. JETP {\bf 34}, 1159 (1972)];
Y. B. Zel'dovich, A. A. Starobinsky, Pis'ma Zh. Eksp. Toer. Fiz. {\bf 26}, 373 (1977) [JETP Lett. {\bf 26}, 252 (1977)]


\bibitem{Ford}
L. H. Ford, Phys. Rev. D {\bf 35}, 2955 (1987)

\bibitem{Turner}
M. S. Turner, Phys. Rev. D {\bf 28}, 1243 (1983)

\bibitem{KKT}
M. Kawasaki, T. Kobayashi, F. Takahashi, Phys. Rev. D {\bf 84} 123506 (2011) [arXiv:1107.6011 [astro-ph.CO]]

\bibitem{Enqvist:2009zf}
  K.~Enqvist, S.~Nurmi, G.~Rigopoulos, O.~Taanila and T.~Takahashi,
JCAP {\bf 0911}, 003 (2009)  [arXiv:0906.3126 [astro-ph.CO]];  
  K.~Enqvist,
Prog.\ Theor.\ Phys.\ Suppl.\  {\bf 190}, 62 (2011)  [arXiv:1012.1711 [astro-ph.CO]]


\bibitem{Tashiro:2003qp}
  H.~Tashiro, T.~Chiba and M.~Sasaki,
Class.\ Quant.\ Grav.\  {\bf 21}, 1761 (2004)  [arXiv:gr-qc/0307068]

\bibitem{EKM}
K.~Enqvist, S.~Kasuya, A.~Mazumdar,
Phys. Rev. Lett. {\bf 93}, 061301 (2004) [arXiv:hep-ph/0311224]

\end{thebibliography}
\end{document}